\documentclass[pra,showpacs,twocolumn,superscriptaddress]{revtex4-1}

\newcommand{\be}{\begin{equation}}
\newcommand{\ee}{\end{equation}}

\usepackage{amsfonts}
\usepackage{amsmath}
\usepackage{graphicx}
\usepackage{amssymb}
\usepackage{amsmath}
\usepackage{amssymb}
\usepackage{graphicx}
\usepackage{color}
\usepackage{epstopdf}
\usepackage{xcolor}
\usepackage[colorlinks=true,allcolors=blue]{hyperref}

\begin{document}
\title{Self-bound droplets of light with orbital angular momentum}

\author{Niclas Westerberg}
\email{nkw2@hw.ac.uk}
\affiliation{SUPA, Institute of Photonics and Quantum Sciences, Heriot-Watt University, Edinburgh EH14 4AS, United Kingdom}
\author{Kali~E. Wilson}
\affiliation{SUPA, Institute of Photonics and Quantum Sciences, Heriot-Watt University, Edinburgh EH14 4AS, United Kingdom}
\author{Callum~W. Duncan}
\affiliation{SUPA, Institute of Photonics and Quantum Sciences, Heriot-Watt University, Edinburgh EH14 4AS, United Kingdom}
\author{Daniele~Faccio}
\affiliation{SUPA, School of Physics \& Astronomy, University of Glasgow, G12 8QQ Glasgow, United Kingdom}
\affiliation{SUPA, Institute of Photonics and Quantum Sciences, Heriot-Watt University, Edinburgh EH14 4AS, United Kingdom}
\affiliation{College of Optical Sciences, University of Arizona, Tucson, Arizona 85721}
\author{Ewan~M. Wright}
\affiliation{College of Optical Sciences, University of Arizona, Tucson, Arizona 85721}
\affiliation{SUPA, Institute of Photonics and Quantum Sciences, Heriot-Watt University, Edinburgh EH14 4AS, United Kingdom}
\author{Patrik \"Ohberg}
\affiliation{SUPA, Institute of Photonics and Quantum Sciences, Heriot-Watt University, Edinburgh EH14 4AS, United Kingdom}
\author{Manuel Valiente}
\email{M.Valiente{\_}Cifuentes@hw.ac.uk}
\affiliation{SUPA, Institute of Photonics and Quantum Sciences, Heriot-Watt University, Edinburgh EH14 4AS, United Kingdom}
\begin{abstract}
Systems with competing attractive and repulsive interactions have a tendency to condense into droplets. This is the case for water in a sink, liquid helium and dipolar atomic gases. Here, we consider a photon fluid which is formed in the transverse plane of a monochromatic laser beam propagating in an attractive (focusing) nonlocal nonlinear medium. In this setting we demonstrate the formation of the optical analogue of matter wave droplets, and study their properties. The system we consider admits droplets that carry orbital angular momentum. We find bound states possessing liquid-like properties, such as bulk pressure and compressibility. { Interestingly, these droplets of light, as opposed to optical vortices, form due to the competition between long-range $s$-wave (monopole) and $d$-wave (quadrupole) interactions as well as diffraction.} 
\end{abstract}

\maketitle
\section{Introduction}
Droplet formation is ubiquitous in nature, its occurrence ranging from classical fluids, such as liquid water in normal conditions, to quantum many-body systems, such as liquid Helium \cite{Dalfavo1995} or atomic mixtures \cite{Pfau1,Pfau2,Tarruell1,Tarruell2017}. In either scenario, the stabilisation of droplets, which are self-bound states, is typically driven by the competition between attractive and repulsive forces between the microscopic constituents of the system. In quantum mechanics, purely attractive forces may still favour droplet formation due to quantum effects. This is the case for zero-range interacting bosonic systems in the universal and few-body limit at zero temperature in two spatial dimensions \cite{HammerSon,bazak2017energy}. Renormalisation effects in the only coupling constant of the system provide the necessary length scale, closely linked to the droplet's size, which in turn provide a mechanism for stabilisation of quantum droplets. More recently, droplets and gas-liquid or gas-droplet transitions in (dipolar) atomic systems have been observed in several groundbreaking experiments \cite{Pfau1,Pfau2,Pfau3,Ferlaino,Tarruell1}. Although the stabilisation mechanism in dipolar atomic systems is due to purely quantum mechanical beyond mean-field effects \cite{Fischer1,Santos,Santos1,Blakie,Pfau4} that require large particle numbers, three- and many-body forces are known to be capable of stabilising droplets in an otherwise collapsing system \cite{Petrov,Bulgac}, and may be the reason for the liquid to Luttinger liquid transition in one-dimensional $^4$He \cite{ValienteOhberg}. { In addition, we should note that the quantum mechanical stabilisation can be modelled by a classical potential in the effective field theory sense, as is done in Ref.~\cite{Pfau2}.}

The most prominent example of natural quantum droplets are large nuclei, some of which have a ground state and part of their excitations well accounted for by the liquid drop model \cite{krane1987nuclear}. Droplet formation is however also present in many other systems, with both local and nonlocal (e.g. power law) interactions \cite{dalfovo2001helium,Bulgac,HammerSon,almand2014quantum,Pfau1,Pfau2,Santos,Santos1,Tarruell1}. In the context of atomic matter waves and nonlinear optics, solitons, and not droplets, are a much more common phenomenon, and have been observed in a variety of scenarios, see Refs.~\cite{burger1999dark,denschlag2000generating,kevrekidis2007emergent,stegeman1999optical} and references therein. Solitons are stationary states that arise in \textit{integrable} systems from the balance between the kinetic energy (i.e. diffraction) and nonlinear interactions. For these to be stable, fine-tuned shapes and densities are required. Droplets, on the other hand, are dynamical objects that can be defined as self-bound, finite-size objects that are stable against perturbations in size, shape and density due to a competition of attractive and repulsive forces. This is the definition we shall use hereafter. 

The connection between droplets and solitons in nonlinear optics has been highlighted by Michinel {\it et al.} \cite{michinel2002liquid,michinel2006turning}, who showed the formal analogy between bosons with competing two- and three-body forces and light in cubic-quintic nonlinear media, forming what they called liquid light. {Moreover, in the context of long-range interactions, similar states have been referred to as \textit{nonlocal solitons} \cite{snyder1997accessible,bang2002collapse,Conti2004Observation,hutsebaut2004single,Xu2005Upper,Malomed1,Malomed2,rotschild2008incoherent,esbensen2012anomalous, zhong2012two,aleksi2012solitons,liang2013spiraling,lu2013power,zhang2014soliton,liang2015spiraling,alberucci2016breather}, \textit{dipole solitons} \cite{lopez2006dipolesoliton,refB1,refB2},  and when rotations are present, \textit{azimuthons} \cite{refB3}.}

In this work, we draw the connection between matter wave droplets and bound states in nonlinear optics with Orbital Angular Momentum (OAM) aiming to explain the underlying mechanisms of the latter. Using the language of atomic quantum fluids, we investigate the properties, underlying mechanisms of formation, and the dynamics of these bound states in detail. We show that they are stable against size and shape perturbations due to a competition between long-range $s$-wave and $d$-wave forces. {Whilst the competing forces are of a different form compared to that of atomic liquids, we nonetheless find that liquid-like properties, such as bulk pressure, compressibility, and a speed of sound can be defined in the system.}

\newpage

The paper is structured as follows: In Section~\ref{sec:analogy} we define the optics-matter wave analogy, followed by Section~\ref{sec:evalH} where the pseudo-energy of bound states are calculated and a new type of expansion for long-range interactions are introduced. In Section~\ref{sec:pwave}, we specifically consider the $p$-wave state, the liquid-like properties of which is explored in Section~\ref{sec:liquid}. The dynamics of a perturbed $p$-wave state is then explored in Section~\ref{sec:dyn}, and concluding remarks are discussed in Section~\ref{sec:conc}.

\section{Optics-matter wave analogy}\label{sec:analogy}
We consider the transverse plane of a monochromatic laser field, for which the photons are effectively massive and two-dimensional. Nonlinearities can be induced by a nonlinear optical medium in such a way that a photon fluid is formed, where superfluidity has also been observed with a repulsive (defocusing) nonlinearity \cite{amo2009superfluidity,amo2011polariton,sanvitto2011all,nardin2011hydrodynamic}. Here the direction of propagation $z$ plays the role of time $t$ in quantum mechanics. We concentrate on nonlocal photon fluids \cite{minovich07experimental,vocke15experimental,vocke2016role}, where the nonlinearity is long-range, formed in the transverse plane of a laser beam propagating in a thermo-optic medium, for which the change in refractive index $\Delta n$ is induced by heat absorption in the medium. Importantly, we are interested in systems with an attractive (focusing) nonlinearity. Superfluid behaviour is thus not expected. As we will show, liquid-like behaviour is {however} present. We shall work with slowly-varying electric field envelopes $E(\mathbf{r},z)$, well described within the paraxial approximation to the wave equation \cite{boyd2003nonlinear},
\begin{equation}
i\frac{\partial E}{\partial z} = -\frac{1}{2k_0}\nabla^2 E-\frac{k_0}{n_0}\Delta n E -\frac{i\alpha}{2}E\equiv \mathcal{H}_*,\label{paraxial}
\end{equation}
where $\nabla^2$ is the Laplacian in the transverse plane ($\mathbf{r}~=~(x,y)$). In Eq.~(\ref{paraxial}), the wave number $k_0$ is given by $k_0~=~2\pi n_0/\lambda$, with $n_0$ the background refractive index of the medium and $\lambda$ the wavelength of the beam, while $\alpha$ is linear absorption coefficient of the medium. The change in refractive index $\Delta n$ is a nonlinear functional of the electric field envelope,
\begin{equation}
\Delta n[E,E^*] = \gamma \int d^2r' R(\mathbf{r}-\mathbf{r}')|E(\mathbf{r}',z)|^2,\label{deltan}
\end{equation}
where $\gamma=\alpha \beta \sigma^2/\kappa$, with $\beta$, $\kappa$ and $\sigma$ the thermo-optic coefficient, thermal conductivity and nonlocal length of the medium (set by the physical size) respectively, and $R(\mathbf{r})$ the medium's thermo-optical response function \cite{vocke15experimental,vocke2016role,roger2016optical}. The response function of the nonlinear medium is well approximated using the distributed loss model, which gives $R(\mathbf{r})=K_0(|\mathbf{r}|/\sigma)/2\pi \sigma^2$ \cite{vocke2016role}, with $K_0$ the modified Bessel function of the second kind of order zero.

The analogy between matter waves and nonlinear optics is drawn by identifying $E$ with the condensate order parameter $\psi$, and $\gamma R$ with the interaction potential $V$. Full analogy with a closed atomic system is achieved for negligible absorption $\alpha$, which is the situation we consider here. { As we shall see, both pseudo-energy $\text{H}_*$ and pseudo-chemical potential $\mu_*$ arise from the conserved quantities of the photon fluid, which will be defined analogously to matter waves. It is easy to see that Eq.~\eqref{paraxial}, neglecting absorption, can be obtained by minimising the following Lagrangian density with respect to $E^*$:
\begin{equation}\label{eq:lagran}
\mathcal{L} = E^*\left(i\partial_z + \frac{\nabla^2}{2k_0} + \frac{k_0}{2n_0}\Delta n\right)E.
\end{equation}
Naturally, due to the $z$-translational invariance of $\mathcal{L}$, the Hamiltonian
\begin{align}\label{eq:hamil}
\text{H}_* &= \int d^2r\; E^*\left(-\frac{\nabla^2}{2k_0}-\frac{k_0}{2n_0}\Delta n\right)E \\ 
& = \int d^2r\; \bigg(\frac{1}{2k_0}\boldsymbol{\nabla} E^*(\mathbf{r},z)\cdot \boldsymbol{\nabla} E(\mathbf{r},z) \nonumber \\
& \quad\quad\quad\quad -  \frac{k_0}{2n_0}\Delta n(\mathbf{r},z) E^*(\mathbf{r},z)E(\mathbf{r},z)\bigg),
\end{align}
is a conserved quantity. We will refer to Eq.~\eqref{eq:hamil} as the pseudo-energy of the photon fluid, in analogy to matter waves. Importantly, another conserved quantity is the power $P$, which will play the role of the number of atoms $N$. That is, 
\begin{equation}\label{eq:power}
P = \int d^2r\; |E(\mathbf{r},z)|^2
\end{equation}
is constant in propagation. We can now define the pseudo-chemical potential $\mu_*$ by minimising the pseudo-energy with Eq.~\eqref{eq:power} as a constraint. In other words, we want to minimise 
\begin{equation}
X[E^*(\mathbf{r},z),E(\mathbf{r},z)] = \text{H}_* - \mu_*\int d^2r\; E^*(\mathbf{r},z)E(\mathbf{r},z).
\end{equation}
Using the pseudo-energy as defined through Eq.~\eqref{eq:hamil}, we find
\begin{equation}
\frac{\delta X}{\delta E^*} = -\frac{1}{2k_0}\nabla^2 E - \frac{k_0}{n_0} \Delta n E - \mu_*E = 0
\end{equation}
and thus we can define the pseudo-chemical potential by
\begin{equation}
\mu_*E = \left[-\frac{1}{2k_0}\nabla^2  - \frac{k_0}{n_0} \Delta n \right] E = \mathcal{H}_*E,
\end{equation}
where $\mathcal{H}_*$ is the pseudo-energy density. This can also be seen as the eigenvalue of the Hamiltonian-density operator $\mathcal{H}_*$, and as such we can make the ansatz $E(\mathbf{r},z) = E(\mathbf{r})e^{-i\mu_* z}$ to obtain the original equation of motion in Eq.~\eqref{paraxial}. From this treatment, it also follows that
\begin{equation}\label{eq:psuedochem}
\mu_* = \frac{\partial \text{H}_*}{\partial P},
\end{equation}
similarly to the chemical potential of a condensate, but where $P \rightarrow N$. Note, this is \textit{not} a chemical potential with respect to the number of photons, but with respect to the power contained in the beam. It represents the amount of pseudo-energy you add to the system by increasing the power by an infinitesimal amount $\delta P$.}

{
\section{Pseudo-energy of bound states}\label{sec:evalH}
We are interested in bound states, and will use an ansatz that generalises the non-rotating results of Hammer and Son \cite{HammerSon}. As we will see, in the process of evaluating the pseudo-energy, we will develop a new type of expansion for highly nonlocal interactions.}

For the ground state we have $E(\mathbf{r},z)=E_p(\mathbf{r})\exp(-i\mu_*z)$, with $\mu_*$ the pseudo-chemical potential. The power-normalised ansatz takes the form
\begin{equation}\label{ansatz}
E_p(\mathbf{r})=\frac{\sqrt{P}}{\sqrt{C_rC_{\phi}}\xi}f\left(r/\xi\right)\Phi(\phi),
\end{equation}
where $\xi$ is a length scale associated with the radial size of the bound state, $f$ is a real-valued radial function, and $\Phi$ encodes the angular dependence. The normalisation constants are $C_r=\int_{0}^{\infty}ds\; s f^2(s)$ and $C_{\phi}=\int_0^{2\pi}d\phi |\Phi(\phi)|^2$. In the following, we work with angular functions $\Phi$ that only contain $\ell = \pm 1$ OAM, that is, $\Phi(\phi)=\exp(i\phi)+\delta \exp(-i\phi)$, with $\delta$ a dimensionless constant. In other words, $\delta$ is the ratio of OAM $\ell = -1$ to OAM $\ell = 1$. {We will now proceed to evaluate the Hamiltonian term-by-term for the bound state ans\"atze of the form~\eqref{ansatz}, as through this we will ultimately find stable shape and size configurations in the usual variational manner. Note that in this, the shape function $f(s)$ is a functional variational parameter.}
{
\subsection{Kinetic pseudo-energy}
Let us start with the kinetic part of the pseudo-Hamiltonian~\eqref{eq:hamil}. We want to calculate the expectation value of the pseudo-energy $\langle \mathcal{H}_{*}\rangle$, using Eq.~\eqref{ansatz} as an ansatz. This yields 
\begin{align}\label{eq:varhamil1}
\text{H}_*^{(1)} &= \frac{P}{2k_0 C_r \xi^2}\left[A_1+\frac{A_2 A_m}{C_\phi}\right] = \frac{P C_1}{2k_0\xi^2},
\end{align}
with dimensionless constants defined as $A_1 = \int_0^\infty ds \; s \left(\frac{df}{ds}\right)^2$, $A_2 = \int_0^\infty ds \;  \left(\frac{f^2}{s}\right)$, $A_m = \int_0^{2\pi} d\phi \; \left|\frac{d\Phi}{d\phi}\right|^2$, and $C_1 = \frac{1}{C_r}\left[A_1+\frac{A_2 A_m}{C_\phi}\right]$. Here $A_1$ and $A_m$ originate from the usual Laplacian of the kinetic Hamiltonian, and $A_2$ accounts for the centrifugal barrier.

\subsection{Interaction pseudo-energy}
As we will see in the following, analytically solving the necessary integrals for the expectation value of the pseudo-energy, in Eq.~\eqref{eq:hamil}, with ans{\"a}tze of the form in Eq.~\eqref{ansatz}, is not possible due to the form of the nonlocal response function $R$. In our case, the bound state is tightly bound (i.e. $\sigma \gg \xi$, see Fig.~\ref{fig:pwaveState}(a)) and the commonly used low-energy (gradient) expansion of $R$ is therefore not appropriate. 

Nonetheless, let us first consider the nonlocal refractive index, Eq.~\eqref{deltan}, in the $\sigma \ll \xi$ limit, in order to gain some intuition. Also, let us denote the intensity, or power density, as $\rho_p(\mathbf{r}) = |E_p(\mathbf{r})|^2$. It follows that 
\begin{align}\label{eq:effrange}
\Delta n(\mathbf{r}) &= \gamma\int \frac{d^2k}{(2\pi)^2}e^{-i\mathbf{k}\cdot\mathbf{r}} R(\mathbf{k})\rho_p(\mathbf{k}) \nonumber\\
&= \gamma\int \frac{d^2k}{(2\pi)^2}e^{-i\mathbf{k}\cdot\mathbf{r}} \frac{\rho_p(\mathbf{k})}{1+\sigma^2 k^2} \nonumber\\
&\simeq \gamma\int \frac{d^2k}{(2\pi)^2} \left[1-\sigma^2k^2\right]e^{-i\mathbf{k}\cdot\mathbf{r}}\rho_p(\mathbf{k}) \nonumber\\
&= \left[\gamma+\gamma\sigma^2\nabla^2_r\right]\rho_p(\mathbf{r}),
\end{align}
where the first step follows from the Fourier transform of the modified Bessel function $K_0$. This is called the effective range expansion in scattering theory \cite{smith2001bose}, or pionless effective field theory in nuclear physics \cite{machleidt2011chiral}, and is valid for nonlocal interaction lengths $\sigma$ much smaller than the characteristic condensate size $\xi$. Physically, this low-energy expansion assumes that the exchange momentum carried by $R(\mathbf{k})$ is much smaller than the condensate momentum $\rho_p(\mathbf{k})$.

However, we are interested in the $\sigma \gg \xi$ regime. We can nonetheless do a similar expansion, which we detail below. We still consider the momentum space picture. As $\sigma \gg \xi$, it follows from line 2 of Eq.~\eqref{eq:effrange} that the exchange momentum $k$ is effectively amplified by the nonlocal length. Therefore, the momentum integral is dominated by $R(\mathbf{k})$, and we may use a low momentum expansion for the photon fluid momentum $\rho_p(\mathbf{k})$. This translates to an effective multipole expansion of the nonlocal refractive index
\begin{widetext}
\begin{align}\label{eq:multipole}
\Delta n(\mathbf{r}) &= \gamma\int \frac{d^2k}{(2\pi)^2}e^{-i\mathbf{k}\cdot\mathbf{r}} R(\mathbf{k})\rho_p(\mathbf{k}) = \int \frac{d^2k}{(2\pi)^2}e^{-i\mathbf{k}\cdot\mathbf{r}} R(\mathbf{k})\int d^2r'\; e^{i\mathbf{k}\cdot\mathbf{r'}}\rho_p(\mathbf{r'}) \nonumber\\
&= \gamma\int d^2r'\; \left[1-x'_{\alpha}\partial_{r}^{\alpha}+\frac{1}{2}x'_{\alpha}x'_{\beta}\partial_{r}^{\alpha}\partial_{r}^{\beta}\right]\rho_p(\mathbf{r'})\int\frac{d^2k}{(2\pi)^2}R(\mathbf{k})e^{-i\mathbf{k}\cdot\mathbf{r}} \nonumber \\ &= \gamma\left[P-d_{\alpha}\partial_{r}^{\alpha}+\frac{1}{2}Q_{\alpha\beta}\partial_{r}^{\alpha}\partial_{r}^{\beta}\right]R(\mathbf{r})
\end{align}
\end{widetext}
We will refer to this expansion as the Long Wavelength Approximation (LWA). Here we expand the plane wave $e^{i\mathbf{k}\cdot\mathbf{r'}}$ to identify the relevant momentum modes that contribute to the nonlocal refractive index. We should note that in order to make an accurate approximation, our reference coordinate system must be aligned with the droplet. Also, a prime implies the primed coordinates, $\partial^\alpha_r = \frac{\partial}{\partial x_\alpha}$, and a (+,+) Einstein summation is implied. For this expansion, we have defined the dipole moment
\begin{equation}
d_\alpha = \int d^2r' \; x'_\alpha \rho_p(\mathbf{r'})
\end{equation}
and the quadrupole moment
\begin{equation}\label{eq:quadrupole}
Q_{\alpha\beta} = \int d^2r' \; x'_\alpha x'_\beta \rho_p(\mathbf{r'}).
\end{equation}

The dipole moment of the ansatz in Eq.~(\ref{ansatz}) vanishes identically ($\mathbf{d}=0$). The quadrupole moment, on the other hand, is non-zero along its diagonal. Substituting the ansatz (\ref{ansatz}) into the definition of the quadrupole moment, we obtain 
\begin{align*}
Q_{\nu,\nu}=\frac{P\xi^2}{C_rC_{\phi}}Q_rq_{\nu,\nu},
\end{align*}
where we have defined $Q_r=\int_{0}^{\infty}ds \; s^3f^2(s)$, $q_{1,1}=\int_0^{2\pi}d\phi \;\cos^2\phi |\Phi(\phi)|^2$, and $q_{2,2}=C_{\phi}-q_{1,1}$. In total, we find that the self-induced refractive index is given by
\begin{align}\label{eq:indexmultipole}
\Delta n(\mathbf{r}) &=\; \gamma P\frac{ K_0(r/\sigma)}{2\pi\sigma^2} + \gamma Q_{12}\sin 2\phi \frac{\;K_2(r/\sigma)}{2\pi\sigma^4} \\ \nonumber 
& + \; \frac{\gamma Q_{11}}{2}\left[ \frac{\cos^2\phi\;K_0(r/\sigma)}{2\pi\sigma^4}+\frac{\cos 2\phi\;K_1(r/\sigma)}{2\pi\sigma^3 r}\right] \\ \nonumber
& + \;  \frac{\gamma Q_{22}}{2}\left[ \frac{\sin^2\phi\;K_0(r/\sigma)}{2\pi\sigma^4}-\frac{\cos 2\phi\;K_1(r/\sigma)}{2\pi\sigma^3 r}\right].
\end{align}
Note that the $d$-wave term ($\propto Q_{ij}$) can take negative values, providing an effective repulsion mechanism, and thus the competing forces that promote stable droplet formation, in stark contrast with the $s$-wave case. 

Finally, we can use the above Eq.~\eqref{eq:indexmultipole} together with our droplet ansatz~\eqref{ansatz} to evaluate the nonlocal interaction part of the Hamiltonian ($\text{H}_*^{(2)}$), yielding 
\begin{widetext}
\begin{align}\label{eq:varhamil2}
\text{H}_*^{(2)} &= -\frac{k_0}{2n_0}\int d^2r\; \Delta n(\mathbf{r})|E_p(\mathbf{r})|^2 \\
&= -\frac{k_0 P \gamma}{2n_0 C_r C_\phi}\Bigg\{\frac{P C_\phi Q_{r}^{K_0}(\xi,\sigma)}{2\pi\sigma^2} + \;\frac{P \xi^2 Q_{r}Q_{r}^{K_0}(\xi,\sigma)}{4\pi\sigma^4 C_{r} C_{\phi}}\left[q_{11}^2+q_{22}^2\right] + \frac{P\xi Q_{r} (Q^{\cos}_{\phi})^2 Q_{r}^{K_1}(\xi,\sigma)}{4\pi\sigma^3 C_{r} C_{\phi}} + \frac{P\xi q_{12} Q_r Q^{\sin}_\phi Q^{K_2}_{r}(\xi,\sigma)}{2\pi\sigma^4 C_r C_\phi}\Bigg\},\nonumber
\end{align}
\end{widetext}
where we have defined the dimensionless constants $Q_{r} = \int_0^\infty ds \; s^3 f^2(s)$, $q_{11} = \int_0^{2\pi} d\phi \; \cos^2\phi \;|\Phi(\phi)|^2$, $q_{22} = \int_0^{2\pi} d\phi \; \sin^2\phi \;|\Phi(\phi)|^2$, $Q^{\cos}_{\phi} = \int_0^{2\pi} d\phi \; \cos 2\phi \;|\Phi(\phi)|^2$, and $Q^{\sin}_{\phi} = \int_0^{2\pi} d\phi \; \sin 2\phi \;|\Phi(\phi)|^2$,
as well as the dimensionless functions $Q_{r}^{K_0}(\xi,\sigma) = \int_0^\infty ds \; s K_0(s\xi/\sigma) f^2(s)$, $Q_{r}^{K_1}(\xi,\sigma) = \int_0^\infty ds \; K_1(s\xi/\sigma) f^2(s)$, and $Q_{r}^{K_2}(\xi,\sigma) = \int_0^\infty ds \; s K_2(s\xi/\sigma) f^2(s)$. We explore the regions of validity for this approximation in Appendix~\ref{app:valid}.
}
\section{Stable configurations}\label{sec:pwave}
\begin{figure*}
\centering
\includegraphics[width=0.6\textwidth]{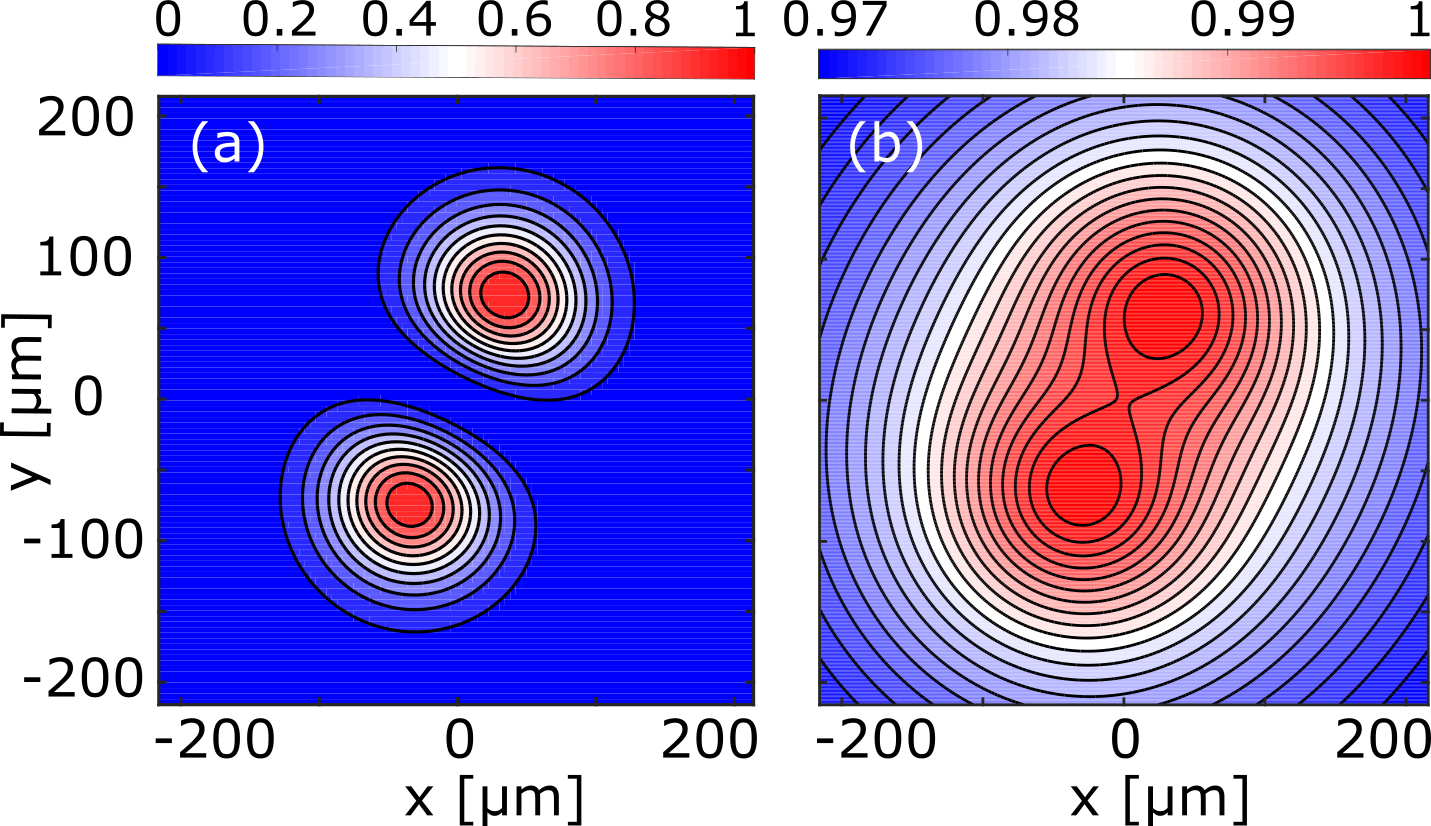}
\caption{
\textbf{(a)} Normalised intensity of a $p$-wave droplet at $P = 1$ W input power and an equal superposition of $\ell = \pm 1$ orbital angular momentum. \textbf{(b)} The corresponding normalised nonlocal interaction potential $\Delta n$.}
\label{fig:pwaveState}
\end{figure*}
{We have now developed a variational expression for the pseudo-energy for bound state ans\"atze. Let us in this section explore the physics further, and discuss stable bound states.} For the case of zero OAM, the ground state of the pseudo-energy functional, after fixing the average power $P=\langle |E|^2\rangle$ (or particle number in the matter wave language), is a single circular bound state. This is well approximated by the model of Snyder and Mitchell \cite{snyder1997accessible} in which the nonlinear refractive index is given by a parabolic function. For reasons that will become apparent, we will refer to this as the $s$-wave bound state. 

{With non-zero OAM however, the ground state is no longer that of the $s$-wave state, and as we shall show in this section, the system instead settles to a symmetry-protected stable state which at a first glance looks like that of two distinct `lobes'.} This state however, truly is a single bound state and we will refer to it as a $p$-wave bound state. Such a state cannot be described within the Snyder-Mitchell framework. Since our system conserves angular momentum, superpositions of $\ell = \pm 1$ modes of the electric field cannot transition into the trivial $s$-wave state in the absence of any perturbations. { In the case of an equal superposition of $\ell = +1$ and $\ell = -1$ modes, which has zero OAM, the $p$-wave state is nonetheless stable as it carries a different symmetry than the $s$-wave state. This state also cannot `fly apart', as the interaction length is much larger than the typical distance between the `lobes'.} Therefore, if these superpositions can find a stable pseudo-energy minimum, they will form bound states that are experimentally feasible to observe. To show that this is the case, we perform a Wick rotation $z\to i\tau$ in the wave equation (\ref{paraxial}), and solve it with initial conditions of the type
$E(\mathbf{r},0)\propto E(r)\left[e^{i\phi}+\delta e^{-i\phi}\right]$. 

In Fig.~\ref{fig:pwaveState}(a) we present the result of imaginary time propagation of Eq.~(\ref{paraxial}) for the intensity distribution with input power $P=1W$, showing that indeed the stable pseudo-ground state configuration is a $p$-wave self-bound state. The formation of the bound state can be attributed to the formation of a double-well-like potential $\Delta n$, as seen in Fig.~\ref{fig:pwaveState}(b). There are two significant features that we immediately infer from Fig.~\ref{fig:pwaveState}(a). Firstly, the zero at the origin is a consequence of the centrifugal barrier $\sim \ell(\ell+1)/r^2$. Secondly, the two lobe structure suggests that the stable configuration has an equal, in absolute value, superposition of $\ell=\pm 1$ modes. Pure $\ell = 1$ or $\ell = -1$ modes would exhibit a ring-like intensity pattern, which we will show is unstable in the following.

\begin{figure*}[]
\centering
\includegraphics[width=0.7\textwidth]{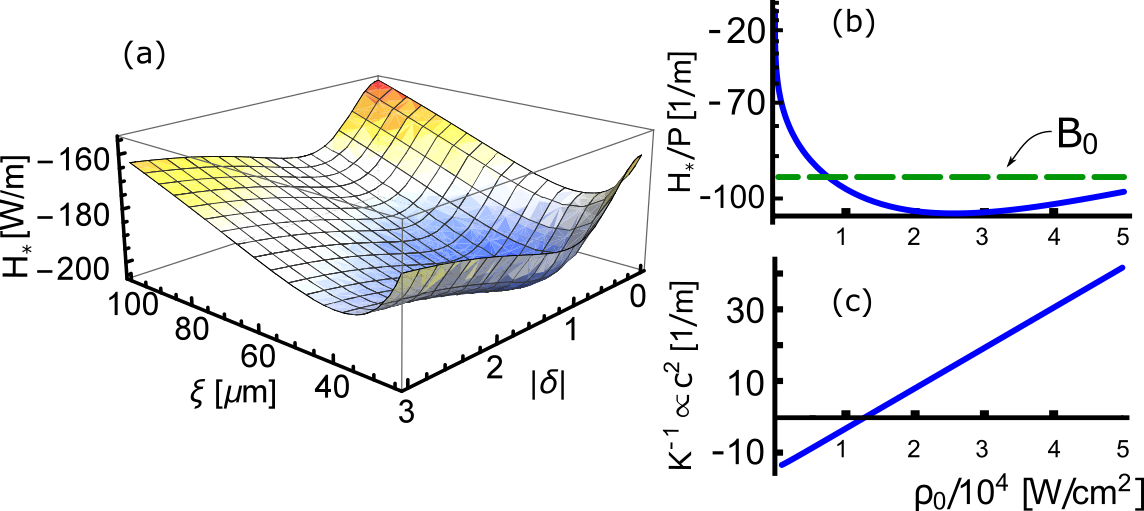}
\caption{
\textbf{(a)} Pseudo-energy surface of the droplet as a function of the droplet size $\xi$ and $|\delta|$ as given by~$\text{H}_*$ for $P = 2$W. Note the global minimum at $|\delta| = 1$ and at the finite size $\xi = \xi_*$. \textbf{(b)} Pseudo-energy per unit power (solid blue line) and binding energy contribution (green dashed line) for $P = 2$W and $|\delta| = 1$ as a function of bulk density (peak power). \textbf{(c)} Corresponding inverse compressibility $K^{-1}$.}
\label{fig:multisurface}
\end{figure*}

We choose $f(s)=s\;e^{-s}$ in Eq.~(\ref{ansatz}), where we have verified that this exponential form gives lower pseudo-energy in the $d$-wave channel than a variety of other forms such as Gaussian functions. In Fig.~\ref{fig:multisurface} we show the energy landscape produced by the ansatz (\ref{ansatz}) as a function of $|\delta|$ and $\xi$. As is clearly observed in Fig.~\ref{fig:multisurface}, the minimum of pseudo-energy occurs at $|\delta| = 1$ (zero net OAM), indicating the formation of a two-peaked intensity pattern, while the ring-shaped, pure $\pm 1$ OAM ($|\delta| = 0$) configuration has the highest energy and is highly unstable. We note here that the Snyder-Mitchell approximation gives the minimum of pseudo-energy for $|\delta|=0$, discussed in Appendix~\ref{sec:SMA}, in stark contrast with numerical simulations and variational calculations within the LWA (see Fig.~\ref{fig:multisurface} and also Ref.~\cite{aleksi2012solitons}). We point out that the $d$-wave quadrupole term is in competition with the $s$-wave monopole term. This is the competition of forces that promotes liquid-like behaviour.

\section{Liquid-like properties}\label{sec:liquid}
{As alluded to earlier, liquid-like features emerge in this system.} The pseudo-Energy per unit Power (EoP), analogous to energy per particle, has a minimum at bulk density (i.e. peak intensity) $\rho_*~=~P/C_r C_\phi \xi_*^2$, where $\xi_*$ is the value of $\xi$ at the pseudo-energy minimum observed in Fig.~\ref{fig:multisurface}(a). This can be seen in Fig.~\ref{fig:multisurface}(b). Expanding around this minimum up to quadratic order in peak intensity $\rho_0~=~P/C_r C_\phi \xi^2$, we find the EoP
\begin{equation}
\frac{\text{H}_*}{P} = -\left|B_0\right| - \frac{3\pi C_\phi}{16 k_0 P}\rho_0 + \frac{9\pi C_\phi^2 n_0}{64 k_0^3 \gamma P^3}\sigma^2 \rho_0^2\label{EoPfunctional}
\end{equation}
where $B_0$ is analogous to the binding energy \cite{Note2} of the bound state at zero density with respect to the $P-1$-particle threshold in the many-particle language. This gives us direct insight into the formation mechanism: Local interactions in the system give rise to the linear term, promoting collapse, which is stabilised by effects related to the nonlocal interaction range ($\sigma$) in the quadratic term. Since our system is dynamical, as opposed to ultracold atomic gases which are cooled to their ground states, the relevant quantity is the EoP, Eq.~(\ref{EoPfunctional}), and not an equation of state \footnote{In thermal equilibrium, and in this context, one would derive the equation of state as the energy per particle \textit{at} the free energy minimum. In this out-of-equilibrium situation, the energy per particle around the minimum plays the equivalent role.}. In other words, our system may be prepared near its pseudo-ground state configuration with a bulk density $\rho_0\ne \rho_*$. The form of the EoP in Eq.~(\ref{EoPfunctional}) at low peak intensity is identical to the form of the equation of state found for {liquid Helium \cite{ValienteOhberg,Dalfavo1995}}, and corresponds to the mean-field approximation with zero-range two-body and three-body forces. It differs only slightly from dipolar BEC:s where $\rho_0^2 \rightarrow \rho_0^{3/2}$ \cite{Note3}.

Furthermore, from the EoP, we can define a (pseudo-) bulk pressure $\mathcal{P}~=~\rho_0^2\;\partial_{\rho_0}[\text{H}_*/P]$ and compressibility $K~=~(\partial_{\rho_0} \mathcal{P})^{-1}$. Zero bulk pressure gives the condition for the bound state energy (minimum). From the compressibility, the inverse of which is plotted in Fig.~\ref{fig:multisurface}(c), we can obtain a spinodal decomposition point \cite{chaikin2000principles}, after which sound waves can propagate in the system. Prior to this point, the system is unstable and droplet nucleation is expected.

We have shown that this $p$-wave bound state not only forms due to a competition of forces, but also attains liquid-like properties. It is thus apparent that it is a droplet, and we shall from here on refer to the state as a $p$-wave droplet in analogy with quantum many-body systems.

\begin{figure*}
\centering
\includegraphics[width=0.7\textwidth]{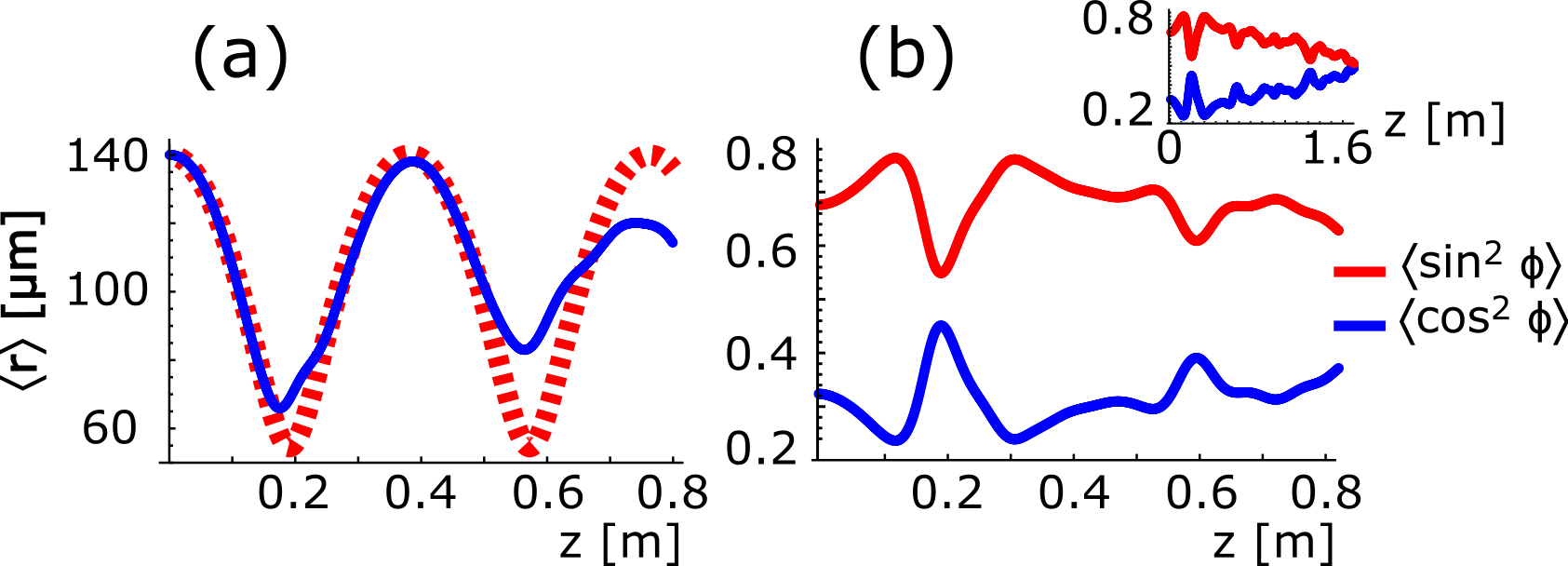}
\caption{
\textbf{(a)} Evolution of the radius $\langle r \rangle$ as predicted by analytical theory (dashed red) and direct numerical simulation (solid blue) with initial conditions $\xi_0 = 70 \mu$m, $|\delta| = 0.9$ and $P = 2$W. \textbf{(b)} Similarly, numerical evolution of $\langle \cos^2\phi \rangle$ and $\langle \sin^2\phi \rangle$ (solid blue and solid red respectively). Droplet rotation can be seen more clearly at longer propagation distance (inset).}
\label{fig:dynamics}
\end{figure*}

\section{Dynamics}\label{sec:dyn}
In order to compare our theoretical variational calculation within the LWA to exact numerical simulations we need to study the ``dynamical" ($z$-dependent) problem. To do so, we first modify the ansatz to account for its non-trivial $z$-dependence. For our dynamical variational parameter $\xi(z)$ (note that by conservation of angular momentum $|\delta|$ is fixed, and the variational analysis is degenerate with respect to the phase of $\delta$), we see that in order to obtain kinetic terms in the Lagrangian $L=\int d^2r \mathcal{L}$ of the form $\sim \partial^2_z \xi$, we need to include a phase term such that the variational ansatz becomes
\begin{equation}
E_p(\mathbf{r},z) \rightarrow E_p(\mathbf{r},z) \exp\left[-\frac{ik_0Z_{\xi}}{2}\int_0^{z}dz'\left[\frac{\mathrm{d}\xi}{\mathrm{d}z'}\right]^2 \right].
\end{equation}
Here we have introduced $z$-independent renormalisation constants, $Z_{\xi}$ and $Z_{\gamma}$, the latter such that $\gamma\to \gamma Z_{\gamma}$. These are necessary since the rate of acquired phase is unknown at this stage, similar to the situation in interacting field theories \cite{srednicki2007quantum,lancaster2014quantum}. Given a number of renormalisation conditions that fix the values of the renormalisation constants, the theory achieves predictive power. Here the constant $Z_\xi$ ($Z_{\gamma}$) can be determined by fixing known linear (nonlinear) effects to either numerical simulations or experimental data. The renormalisation constants manifest themselves as counterterms in the renormalised Lagrangian, which reads
\begin{align}
L =\;\; &\frac{Pk_0}{2}\left(\frac{\mathrm{d}\xi}{\mathrm{d}z}\right)^2 - \text{H}_* \nonumber\\ 
&+\; (Z_{\xi}-1)\frac{Pk_0}{2}\left(\frac{\mathrm{d}\xi}{\mathrm{d}z}\right)^2 -(Z_{\gamma}-1)\text{H}_*^{(2)},
\end{align} 
where $\text{H}_*^{(2)}$ is the part of the Hamiltonian containing only the nonlinearity introduced by $\Delta n$ in Eq.~\eqref{deltan}. By minimising the Lagrangian, we find the equation~of~motion
\begin{equation}\label{eq:xiDyn}
\frac{\mathrm{d}^2\xi}{\mathrm{d}z^2}=-\frac{1}{Pk_0Z_{\xi}}\frac{\partial \text{H}_*}{\partial \xi}.
\end{equation}
Upon renormalisation, we find the constants to have values $Z_{\xi}=4$ and $Z_{\gamma}=0.76$. { These are fixed by the initial propagation of a single low $P$ simulation, and a single high $P$ simulation, respectively for $Z_{\xi}$ and $Z_\gamma$. Importantly, we can now use these values for \textit{any} $P$, $\xi$, or $\delta$.}

\begin{figure*}[]
\centering
\includegraphics[width=0.7\textwidth]{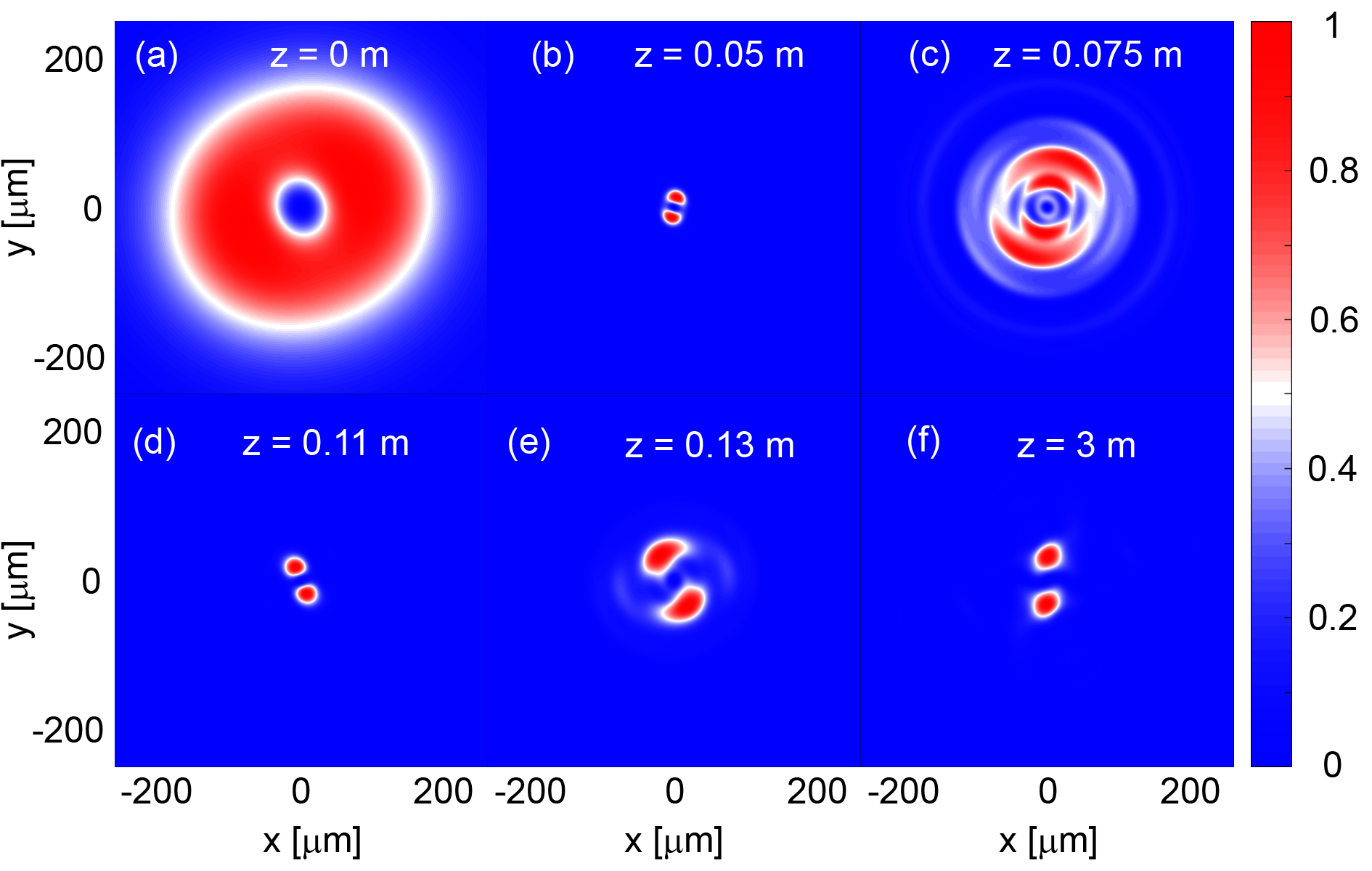}
\caption{
\textbf{(a)-(f)} Evolution snapshots for an initial power $P = 15$W, radial size $\xi_0 = 70 \mu$m and $|\delta| = 0.05$. Observe the initial collapse in (b), followed by a violent rebound in (c). This collapse-rebound cycle continues (less violently) until the system has re-distributed most of its pseudo-energy into $p$-wave droplet binding energy, such as in (f). Excess pseudo-energy emerges as surface vibrations, see Supplementary Video 2.}
\label{fig:collapse}
\end{figure*}

From Fig.~\ref{fig:multisurface}, we can expect dynamics in the form of oscillations, if the input electric field envelope has $\xi$ or $\delta$ slightly away from the minimum. The frequency of these oscillations can be related to the surface tension of the droplet \cite{michinel2002liquid,Note4}. We choose as initial conditions $\xi_0=70\mu$m and $\delta=0.9$ with input power $P=2$W and evolve using a split-step propagation algorithm \cite{roger2016optical}. In Fig.~\ref{fig:dynamics}, we show the evolution of the radius $\langle r \rangle$, $\langle \sin^2\phi \rangle$ and $\langle \cos^2\phi \rangle$ at different~$z$. We observe oscillations in the radius as well as an overall rotation. The variational model is in excellent qualitative, and good quantitative agreement with the exact dynamics of the system. In particular, the main feature of the evolution, i.e. oscillations in size, are properly reproduced by our model, including the correct period of the oscillations. If we further consider the angular dynamics, we see that the overall trend in Fig~\ref{fig:dynamics}(b) implies that the droplet is rotating, and the small oscillations indicate that the angular distribution oscillates (for more details of this rotation, see Supplementary Video 1). The latter can be attributed to the system attempting to reach the $|\delta| = 1$ minimum, but due to conservation of angular momentum is only able to temporarily scatter momenta away from the bound state.

We shall now consider the far-from-equilibrium situation, see Fig.~\ref{fig:collapse} (see also Supplementary Video 2). Here the initial condition, seen in Fig.~\ref{fig:collapse}(a), is a slightly perturbed vortex ($|\delta| = 0.05)$ with input power $P=15$W. This is in the unstable regime (left of the spinodal decomposition point, i.e. where $K^{-1} \leq 0$) and droplet nucleation is expected. Indeed, the $p$-wave droplet emerges after a few violent collapse-rebound cycles. Whilst most of the pseudo-energy is either shed or re-distributed into binding energy, some excess pseudo-energy manifests itself as surface vibrations. 

\section{Conclusions}\label{sec:conc}
In conclusion, we have found and described bound states carrying non-zero orbital angular momentum in a photon fluid with nonlocal attractive (focusing) interactions (nonlinearity). In particular, using matter-wave analogies and developing theoretical tools not typically used in optics, we found that these states are stabilised by a competition of long-range $s$-wave and $d$-wave forces, exhibit liquid behaviour, and are thus a type of \textit{droplet}. In fact, these bound states are similar to the droplets recently found experimentally in dipolar atomic gases, albeit with a different stabilisation mechanism. The observed rotation may be linked to self-induced synthetic magnetic fields recently introduced in photon fluids \cite{Synthetic2016Westerberg}. 

\appendix
{
\section{Validity of the LWA}\label{app:valid}
A commonly used expansion for a nonlocal refractive index is the so-called Snyder-Mitchell approximation, where the refractive index is simply proposed to take the form of a parabola \cite{snyder1997accessible}. This would correspond to only keeping the $s$-wave term of the above expansion, and further approximating the medium's response function $R(\mathbf{r})\propto r^2$. Let us now examine the regions of validity of the Long Wavelength Approximation, and compare this to the Snyder-Mitchell approach. Note that, we will show that the Snyder-Mitchell approach is incorrect for non-zero OAM in Appendix~\ref{sec:SMA}. We start from Eq.~\eqref{deltan} which gives the thermally-induced refractive-index change from the distributed loss model as
\begin{equation}
\Delta n({\bf r},z) = \gamma \int_{  }^{  } d^2r' R({\bf r}-{\bf r}') |E({\bf r}',z)|^2  , 
\end{equation}
with $|E({\bf r},z)|^2$ scaled such that it yields the transverse intensity profile. As an illustrative example we use the case of a spherically symmetric ($s$-wave) Gaussian of power $P$ and spot size $\xi$
\begin{equation}
|E({\bf r},z)|^2 = {2P\over \pi \xi^2} e^{-2r^2/\xi^2}  .
\end{equation}
For such a symmetric intensity profile the refractive-index profile may be recast in cylindrical coordinates as
\begin{equation}\label{DnGreen}
\Delta n(r,z) = \gamma \int_{0}^{\infty} |E(\zeta ,z)|^2 G_0(r,\zeta;a) \zeta d\zeta , 
\end{equation}
where the Green's function is given by Eq. (7) of Ref.~\cite{Yak05}, but where in their notation $\zeta \rightarrow \xi$ and $W$ denotes the spot size instead of $\xi$. In our case, the parameter $a$ is equal to the nonlocal length $\sigma$.

We are interested in the case when the Gaussian spot size is much smaller than the nonlocal length $\xi\ll\sigma$, and we set $\xi=0.005\sigma$ as an example. The solid line in Fig.~\ref{fig:scaledDNcomapring}(a) shows the scaled index change $\Delta n/\gamma P$ versus $r/\sigma$ calculated numerically according to Eq.~\eqref{DnGreen} above, and we shall use this example as a test bed for the approximations employed in the paper.

The long wavelength approximation is given by Eq.~\eqref{eq:indexmultipole}, and this yields 
\begin{equation}\label{DnLWA}
{\Delta n\over \gamma P}  = {K_0(r/\sigma)\over 2\pi\sigma^2}\left[1+\frac{1}{2}\left(\frac{\xi}{\sigma}\right)^2\right],
\end{equation}
We note that this approximation for the scaled index change $\Delta n/\gamma P$ does not involve the Gaussian spot size. This arises since the narrow Gaussian with $\xi\ll\sigma$ acts like a $\delta$-function multiplied by the power $P$. The dashed line in Fig.~\ref{fig:scaledDNcomapring}(a) shows the scaled index change $\Delta n/\gamma P$ versus $r/\sigma$ according to the long wavelength approximation Eq.~\eqref{DnLWA}. We see as expected that the long wavelength approximation does not work well near the origin but it improves at larger distances. For the $p$-wave case discussed in the text, the centrifugal barrier makes the droplets avoid short distances, as we have seen, and short distances become irrelevant.

\begin{figure}[]
\centering     
\includegraphics[width=0.45\textwidth]{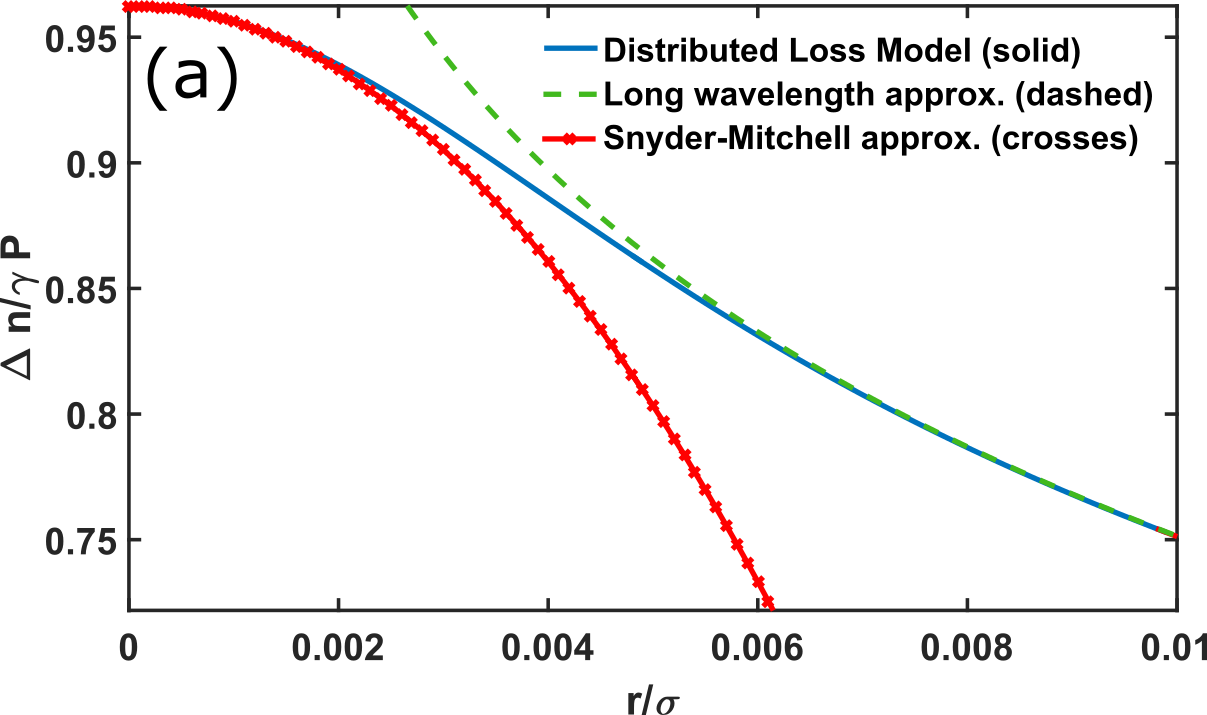}
\includegraphics[width=0.441\textwidth]{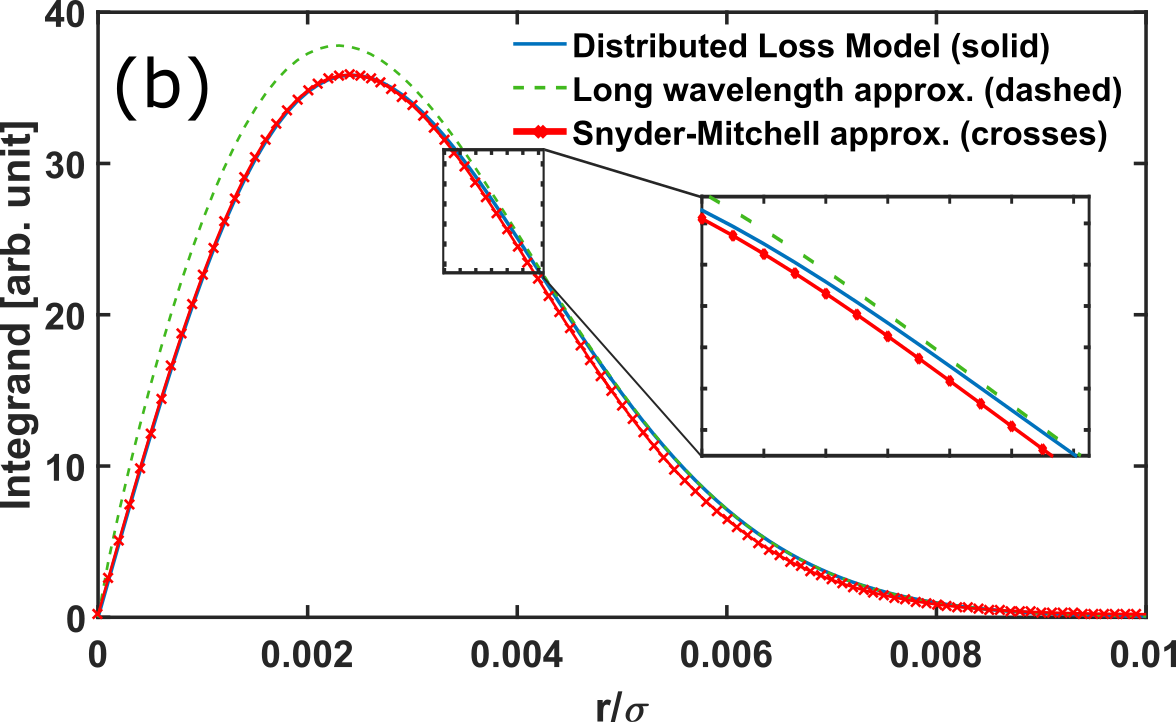}
\caption{\textbf{(a)} Scaled index change $\Delta n/\gamma P$ versus $r/\sigma$ according to the long wavelength approximation Eq.~\eqref{DnLWA} (dashed), according to the Snyder-Mitchell approximation Eq.~\eqref{DnSMA} (solid plus crosses) and the exact index change (solid). \textbf{(b)} The exact solution for $\Delta n(r)$ (solid), long wavelength approximation (dashed), and Snyder-Mitchell approximation (solid plus crosses). Notice that the different approximate refractive-index profiles both give similar integrands. One should note that a $p$-wave droplet is expected to be localised at radii $r\gtrsim 0.0035 \sigma$, where the long wavelength approximation better approximates the distributed loss model. Inset shows the crossover region.}
\label{fig:scaledDNcomapring}
\end{figure}

We will discuss the Snyder-Mitchell model in more detail in Appendix~\ref{sec:SMA}, but in short the thermally-induced refractive-index change is given by
\begin{equation}\label{DnSMA}
\Delta n(r) - \Delta n(0)  = -{\gamma P\over 4\sigma^2 A_{eff}} r^2  , 
\end{equation}
with $A_{eff}=\pi \xi^2/2$ for the Gaussian example, and $\Delta n(0)$ is the on-axis index change. The solid line with crosses in Fig.~\ref{fig:scaledDNcomapring}(a) shows the scaled index change $\Delta n/\gamma P$ versus $r/\sigma$ according to the Snyder-Mitchell approximation, Eq.~\eqref{DnSMA}. In contrast to the long wavelength approximation, the Snyder-Mitchell approximation works very well close to the origin but deviates at larger distances. In particular, the Snyder-Mitchell approximation provides a better approximation to the index profile for distances $r\lesssim \xi/\sqrt{2} = 0.0035\sigma$, that is over the spatial extent of the Gaussian example of width $\xi$. Whilst the on-axis Gaussian provides an illustrative example, we should note that $p$-wave droplet is localised off-axis. In fact, we expect the peaks and thus the bulk of the field intensity to be at radii $r\gtrsim \xi/\sqrt{2}$ and thus the long wavelength approximation is preferred.

An apparent issue arises in that we are using noticeably different forms of the scaled index change $\Delta n/\gamma P$ from the long wavelength approximation and the Snyder-Mitchell but hoping to address the same physics. The question is how useful information may be obtained based on the long wavelength approximation which actually diverges near the origin. The answer lies in the fact that we use the LWA in a variational (Lagrangian or Hamiltonian) calculation that involves the integral of the refractive index profile and the intensity of the form
\begin{equation}
\int_0^\infty \underbrace{|E(r)|^2 \Delta n(r) r}_{integrand} dr.
\end{equation}

In Fig.~\ref{fig:scaledDNcomapring}(b) we plot the underbraced integrand versus $r/\sigma$ from this integral for the exact solution for $\Delta n(r)$ (solid), long wavelength approximation (dashed), and Snyder-Mitchell approximation (solid plus crosses). Here we see that despite the differences in the approximate refractive-index profiles they give very similar results for the integrand. Thus from the perspective of the variational methods the different approaches should yield similar results in the spherically symmetric case since they depend on integrals as above.

\section{Comparison to Snyder-Mitchell approximation}\label{sec:SMA}
\begin{figure}[]
\centering
\includegraphics[width=0.25\textwidth]{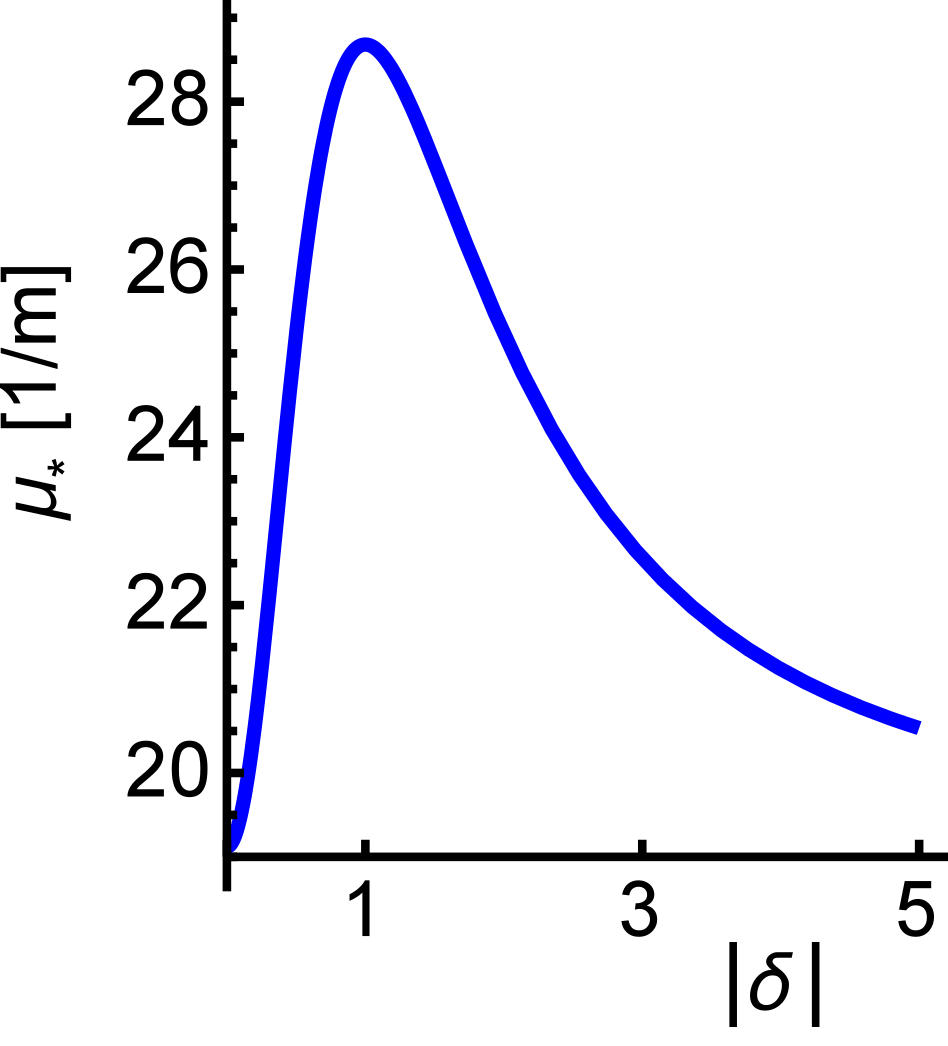}
\caption{
Pseudo-chemical potential surface of the droplet as a function of the variational parameter $\delta$ as given by Eq.~\eqref{eq:psuedoEnergy} from the Snyder-Mitchell model with $P = 1$W. As can be seen, this energy landscape is qualitatively different from the one seen in Fig.~2(a). This is also qualitatively different from numerical evidence.}
\label{fig:SMdeltasurface}
\end{figure}

We shall here explore the $p$-wave droplet characteristics under the assumption that the Snyder-Mitchell model of the nonlocal response function is correct. There are a plethora of Snyder-Mitchell models, many of which assume a constant harmonic frequency \cite{snyder1997accessible,alberucci2016breather,Conti2004Observation,alberucci2015spatial,zhong2012two,bang2002collapse}. These models are azimuthally symmetric and droplet-shape independent so it comes as no surprise that the pseudo-energy is $\delta$-independent. However, we can improve upon this by taking the shape into account. As before, let us start at the paraxial wave equation:
\begin{equation}\label{eq:paraxial1}
i\frac{\partial E(\mathbf{r},z)}{\partial z} = -\frac{1}{2k_0}\nabla^2 E(\mathbf{r},z) - \frac{k_0}{n_0} \Delta n(\mathbf{r},z) E(\mathbf{r},z).
\end{equation}
In the Snyder-Mitchell regime, the nonlocal response of the medium is approximated as $\Delta n(\mathbf{r}) \simeq -\Omega^2r^2$, where $\Omega$ is the harmonic oscillator frequency given by the relation
\begin{equation}
\Omega^2 = \frac{\alpha\beta P}{4\kappa A_{\text{eff}}} = \frac{\gamma P}{4\sigma^2 A_{\text{eff}}}
\end{equation}
and where
\begin{equation}\label{eq:effarea}
A_{\text{eff}} = \frac{\left(\int d^2r |E(\mathbf{r})|^2\right)^2}{\int d^2r |E(\mathbf{r})|^4}
\end{equation}
is the effective area of the beam. We are looking for droplets with orbital angular momentum $\ell \neq 0$, and thus we make the ansatz
\begin{equation}\label{eq:ansatz}
E(\mathbf{r},z) = E_d(\mathbf{r})e^{i\ell\phi}e^{-i\mu_*z}.
\end{equation}
This has solutions of the form
\begin{equation}
E_p(r,\phi) = r^{|\zeta|}{}_1F_1(-n,|\zeta|+1,2r^2/\xi^2)e^{-r^2/\xi^2}e^{i m\phi}
\end{equation}
where $m$ is an integer, $n$ is a positive integer and $|\zeta| = |m+\ell|$ and ${}_1F_1$ is the confluent hypergeometric function of the first kind. The pseudo-chemical potential of this family of solutions is given by
\begin{equation}\label{eq:psuedoEnergy}
\mu_*=\frac{k_0\gamma P}{4n_0 C_{\text{area}} \sigma^2}\left(2n+|\zeta|+1\right).
\end{equation}
Here we have defined $C_{\text{area}}\equiv A_{\text{eff}}/\xi^2$. Focusing on the case of $n=0$, the normalised droplet takes the form
\begin{align*}
E_d(r,\phi) &= \frac{2^{\frac{|\zeta| +1}{2}} \sqrt{P} \xi ^{\frac{1}{2} (-2 |\zeta| -2)}}{\sqrt{\pi } \sqrt{\Gamma (|\zeta| +1)}}\times \\
& \quad\quad\quad r^{|\zeta|}{}_1F_1(0,|\zeta|+1,2r^2/\xi^2) e^{{-r^2/\xi^2}}e^{{\pm i m\phi}}.
\end{align*}
For this to be a self-consistent solution, we require that 
\begin{equation}
\frac{4}{\xi^4} = \frac{2k_0^2\Omega^2(\xi)}{n_0} 
\end{equation}
and thus 
\begin{equation}
\frac{4}{\xi^4} = \frac{2k_0^2\gamma P}{4\sigma^2 A_{\text{eff}}(\xi)} = \frac{2k_0^2\gamma P}{4n_0\sigma^2 C_{\text{area}}\xi^2}
\end{equation}
where in the latter step we used Eq.~\eqref{eq:effarea} to calculate the effective area. In the above, 
\begin{equation}\label{eq:Carea}
C_{\text{area}} = \frac{ 4^{|\zeta| } \Gamma (|\zeta|+1)^2\pi}{\Gamma (2 |\zeta| +1)}.
\end{equation}
Solving the self-consistency relation for the characteristic size $\xi$ yields 
\begin{equation}
\xi = \sqrt{\frac{8C_{\text{area}}n_0}{k_0^2\gamma P}}\sigma.
\end{equation}
So far, this looks quite far from the observed $p$-wave droplet. However, let us look closer at the pseudo-chemical potential $\mu_*$. If we let $\ell = 1$, then we notice that $m = 0$ and $m = -2$ both have $|\zeta| = 1$. Therefore the $m= 0$ and $m = -2$ state has the same pseudo-energy. In other words, the ground state is degenerate. We can now form a superposition of $s$- and $d$-wave states (for OAM $\ell=1$) as a superposition of the two degenerate ground states, i.e. the electric field takes the form
\begin{equation}\label{eq:SMpwave}
E^{\text{p-wave}}_d(r,\phi)= \frac{2 \sqrt{P} }{\sqrt{\pi } \sqrt{1+\delta^2} \xi^2}r e^{-\frac{r^2}{\xi ^2}}\left[1+\delta  e^{-2 i \phi }\right].
\end{equation}
Here we should point out that in general, a superposition has a different effective area than either of its constituents, that is, the area is not given by Eq.~\eqref{eq:Carea} in general, and thus the pseudo-chemical potential will vary with $\delta$. The pseudo-chemical potential and the pseudo-energy is connected through Eq.~\eqref{eq:psuedochem}, therefore the landscape of the pseudo-chemical potential maps directly to the pseudo-energy surface. In Fig.~\ref{fig:SMdeltasurface} we see that the Snyder-Mitchell model predicts qualitatively different behaviour to the low wavelength approximation, c.f. Fig.~\ref{fig:multisurface}. The long wavelength approximation is in agreement with numerical results, we must conclude that the Snyder-Mitchell model is not sufficient to capture the physics addressed in this work.
}

\begin{acknowledgments}
The authors would like to acknowledge insightful discussions with Calum Maitland, Magnus Borgh and Luis Santos. NW and CWD acknowledges support from EPSRC CM-CDT Grant No. EP/L015110/1. P\"O and MV acknowledge support from EPSRC grant No. EP/M024636/1. D.F. acknowledges financial support from the European Research Council under the European Unions Seventh Framework Programme Grant No. (FP/2007-17172013)/ERC GA 306559 and EPSRC (U.K., Grant No. EP/J00443X/1).
\end{acknowledgments}

\bibliographystyle{apsrev4-1}

\end{document}